\documentclass[aps,prb,reprint,amsmath,amssymb,,superscriptaddress,floatfix,floats]{revtex4-2}
    \usepackage{graphicx}           
    \usepackage{dcolumn}            
    \usepackage{bm}                 
    \usepackage{hyperref}           
    \hypersetup{colorlinks=true, linkcolor=blue, citecolor=blue, urlcolor=blue, pdftitle={Metallic layered materials with magnetic frustration: an ARPES view of the LnTAl4X2 family}, pdfauthor={PedroRezende}}
    \usepackage{orcidlink}
    \usepackage{color,soul}
    \usepackage{lipsum}
    \usepackage{xcolor}
    
\begin{document}
\preprint{APS/123-QED}

\title{Metallic layered materials with magnetic frustration: An ARPES view of the SmAuAl\texorpdfstring{$_4$}{4}Ge\texorpdfstring{$_2$}{2} and TbAuAl\texorpdfstring{$_4$}{4}Ge\texorpdfstring{$_2$}{2}}

\thanks{These authors contributed equally to this work}
\author{P. Rezende-Gon\c calves$^*$}
    \email{pedrohrg@ufmg.br}
    \affiliation{Universit\'{e} Paris-Saclay, CNRS, Institut des Sciences Mol\'{e}culaires d'Orsay, 91405 Orsay, France}
    \affiliation{Departamento de F\'{i}sica, Universidade Federal de Minas Gerais, Av. Pres. Antonio Carlos, 6627 Belo Horizonte, Brazil}
\author{A. Antezak$^*$}
    \email{alexandre.antezak@universite-paris-saclay.fr}
    \affiliation{Universit\'{e} Paris-Saclay, CNRS, Institut des Sciences Mol\'{e}culaires d'Orsay, 91405 Orsay, France}
\author{T. Kato}
    \affiliation{Universit\'{e} Paris-Saclay, CNRS, Institut des Sciences Mol\'{e}culaires d'Orsay, 91405 Orsay, France}
    \affiliation{Advanced Institute for Materials Research (WPI-AIMR), Tohoku University, Sendai 980-8577, Japan}
\author{K. Feng}
    \affiliation{Department of Physics, University of California, San Diego, California 92093, USA}
\author{F. Fortuna}
    \affiliation{Universit\'{e} Paris-Saclay, CNRS, Institut des Sciences Mol\'{e}culaires d'Orsay, 91405 Orsay, France}
\author{P. Le F\`{e}vre}
    \affiliation{Synchrotron SOLEIL, L'Orme des Merisiers, Saint-Aubin-BP48, 91192 Gif-sur-Yvette, France}
    \affiliation{Univ Rennes, CNRS, IPR - UMR 6251, F-35000 Rennes, France}
\author{M. Rosmus}
\author{N. Olszowska}
\author{T. Sobol}
    \affiliation{National Synchrotron Radiation Centre SOLARIS, Jagiellonian University, Czerwone Maki 98, 30-392 Krak\'{o}w, Poland}
\author{D. J. Singh}
    \affiliation{Department of Physics and Astronomy, University of Missouri, Missouri 65211, USA}  
\author{R. E. Baumbach}
    \affiliation{Physics Department, University of California, Santa Cruz, California 95064, USA}
\author{A. F. Santander-Syro}
\author{E. Frantzeskakis}
    \email{emmanouil.frantzeskakis@universite-paris-saclay.fr}
    \affiliation{Universit\'{e} Paris-Saclay, CNRS, Institut des Sciences Mol\'{e}culaires d'Orsay, 91405 Orsay, France}

\date{\today}

\begin{abstract}
Compounds of the new materials class \textit{LnT}Al$_4$\textit{X}$_2$ (\textit{Ln} = lanthanide, \textit{X} = tetrel, \textit{T} = transition metal) host exotic magnetic phenomena due to geometric frustration induced by their triangular lattice. Complex spin arrangements, magnetic fluctuations and double magnetic transitions have been well observed by means of magneto-transport. Nevertheless, the experimental electronic structure of this family of materials has been poorly studied. We have investigated the experimental electronic structure of two members of this class of materials: SmAuAl$_4$Ge$_2$ and TbAuAl$_4$Ge$_2$. By means of Angle-Resolved PhotoEmission Spectroscopy (ARPES) accompanied by Density Functional Theory calculations (DFT), we reveal common trends and features, the important effect of localized spin moments on the electronic structure, the presence of surface-localized electronic states and the nature of the surface termination layer. Low-dimensionality, exchange interaction, and spin-orbit coupling are all important ingredients of the electronic structure.
\end{abstract}

\maketitle

\section{Introduction}
Scientific interest on the novel series of metallic rare earth lanthanides with the chemical formula \textit{LnT}Al$_4$\textit{X}$_2$ (\textit{Ln} = lanthanide, \textit{X} = tetrel, \textit{T} = transition metal) is inherent to their chemical composition and structure. On the one hand, their chemical composition favors strong electron correlations via the presence of elements with partially-filled \textit{f} shells. The latter act as magnetic impurities with localized spins that interact with the surrounding cloud of conduction electrons. The coupling of the localized spins to the conduction electrons (Kondo interaction \cite{kondo_resistance_1964}) and the indirect exchange interaction of the localized spins themselves (RKKY interaction \cite{ruderman_indirect_1954}) act together to generate the driving force behind many phenomena in strongly electron correlated systems \cite{pruser_interplay_2014,iglesias_revisited_1997}. Such phenomena include charge and spin order, unconventional superconductivity, heavy fermions, intermediate valences, complex magnetism, etc \cite{alexandradinata_future_2022}.

On top of this rich playground of physics that goes beyond the well-established model of non-interacting electrons, the crystal structure of the \textit{LnT}Al$_4$\textit{X}$_2$ compounds conspires to add magnetic frustration into the picture. This is the phenomenon where the atomic arrangement induces frustration in the alignment of the spin magnetic moments and may lead to complex magnetic phenomena related to collective behaviour such as quantum spin liquids and non-collinear spin configurations \cite{balents_spin_2010,coleman_frustration_2010}. The concept of geometric frustration was first introduced in the context of triangular lattices, where no stable spin arrangement can minimize the total energy of the crystal \cite{wannier_antiferromagnetism_1950}. Some examples of geometrically frustrated systems are the Kagome lattice \cite{PhysRevResearch.2.012045,lu_observation_2022,norman_colloquium_2016}, the pyrochlore lattice \cite{taguchi_spin_2001,taguchi_magnetic_2003} and the spin-ice materials \cite{castelnovo_magnetic_2008,wang_artificial_2006}. Since strongly-correlated electron systems, and systems with magnetic frustration are extremely interesting on their own, one can easily imagine the impact of a new series of compounds that gives access to both concepts and underlying phenomena. Examples of \textit{f}-electron materials with magnetic frustration have been already reported and include CePdAl \cite{fritsch_approaching_2014}, YbAgGe \cite{sengupta_geometrical_2010} and CeRhSn \cite{tokiwa_characteristic_2015}. All of these compounds present complex magnetic behavior which can be attributed to either the interaction of the \textit{f}-electrons with the conduction states or the crystal structure itself. The difficulty in underpinning the underlying mechanism is the presence of strong hybridization between the localized  \textit{f}-states and the conduction band. Here lies the novelty of the \textit{LnT}Al$_4$\textit{X}$_2$ series: it is a materials family that combines geometric frustration and the presence of magnetic moments, while the hybridization strength between localized and itinerant electrons remains weak.

Magnetic frustration manifests itself in exotic physics in the \textit{LnT}Al$_4$\textit{X}$_2$ compounds as complex magnetic behavior and multiple magnetic transitions are the rule. NdAuAl$_4$Ge$_2$ undergoes a double transition to a ground state with long-range magnetic order at zero field, but its field-induced metamagnetic transitions are poorly understood \cite{PhysRevMaterials.7.024423}. CeAuAl$_4$Ge$_2$ presents magnetic fluctuations near 15~K and ordering of magnetic moments below 1.4~K, with the details of the ordering still unknown \cite{zhang_electronic_2017}. GdNiAl$_4$Ge$_2$ is a candidate for hosting non-trivial spin textures because it presents history-dependent spin structures with complex relaxation dynamics \cite{PhysRevMaterials.7.124409}. Finally, GdAuAl$_4$Ge$_2$, TbAuAl$_4$Ge$_2$ and SmAuAl$_4$Ge$_2$ present a double transition at higher ordering temperatures with very complex ordering of the magnetic moments in their ground states \cite{PhysRevB.109.014436,feng_magnetic_2022,leahy_field-induced_2022}.

The aforementioned physics phenomena in the compounds of the \textit{LnT}Al$_4$\textit{X}$_2$ family have been experimentally investigated by means of magneto-transport. On the other hand, their electronic structure -although still poorly studied- might give new insights on their complex behavior. For example, its experimental determination may answer whether the \textit{f}-electrons can be safely ignored due to their weak hybridization with the conduction states or there are different ways of how they (or the presence of localized spins) manifest themselves on the Fermi surface. Moreover, the experimental electronic structure may reveal the presence of surface-localized electronic states that can influence the materials’ properties. In any case, a combined view of the experimental electronic structure of the \textit{LnT}Al$_4$\textit{X}$_2$ compounds is expected to pinpoint certain trends and features that might contribute in explaining their exotic behavior. Up to this date, there is however only a single Angle-Resolved PhotoEmission (ARPES) work on a compound from this materials family \cite{zhang_multiple_2023}. In the present work, we aim to fill this gap by performing a detailed ARPES study on TbAuAl$_4$Ge$_2$ and SmAuAl$_4$Ge$_2$, accompanied by Density Functional Theory (DFT) calculations. Our goal is to reveal the experimental electronic structure of the two compounds focusing on their similarities and on a comparison to their non-magnetic analogue YAuAl$_4$Ge$_2$. On one hand, our data show that the non-magnetic YAuAl$_4$Ge$_2$ can capture the main experimental findings of the electronic structure in a satisfactory way, thereby experimentally confirming that there is negligible hybridization between the \textit{f}-electrons and the conduction states. On the other hand, there are certain details of the electronic structure near the Fermi level for which the magnetic moments of the \textit{f}-electrons cannot be ignored showing that they contribute -indirectly- to the features of the Fermi surface. Finally, we present experimental evidence that a full description of the electronic structure needs to take into account surface-localized features and the nature of the surface termination layer.

\section{Materials and Methods}

\subsection{Sample Growth:}
(Sm/Tb)AuAl$_4$Ge$_2$ single crystals were grown using a molten metal flux following the procedure detailed in Refs. \cite{wu_reaual4ge2_2005,PhysRevB.109.014436,feng_magnetic_2022}. Elements with purities $>$ 99.9\% were combined in the molar ratio 1(\textit{Ln}):1(Au):10(Al):5(Ge) and loaded into 2~mL alumina Canfield crucibles. The crucibles were sealed under vacuum in quartz tubes, heated to 1000~$^{\circ}$C at a rate of 83~$^{\circ}$C/h, kept at 1000 ~$^{\circ}$C for 15~h, then cooled to 860~$^{\circ}$C at a rate of 7~$^{\circ}$C/h. The melt was then annealed at 860~$^{\circ}$C for 48~h with the goal of improving sample quality and surfaces. In order to minimize thermal shock, this was followed by cooling down to 700~$^{\circ}$C at a rate of 12~$^{\circ}$C/h, after which excess flux was removed by centrifuging the tubes at 700~$^{\circ}$C. Crystals typically form as three-dimensional clusters, where individual crystals with dimensions on the order of 2~mm and hexagonal or triangular facets associated with the ab plane can be isolated. These crystals were then cleaved under vacuum in preparation for ARPES measurements.

\subsection{Experimental Methods:}
ARPES measurements were carried out using hemispherical electron analyzers with vertical slits at three different beamlines of Synchrotron facilities: the CASSIOPEE beamline of Synchrotron SOLEIL (France) and the URANOS and PHELIX beamlines of Synchrotron SOLARIS (Poland). In order to generate pristine surfaces, SmAuAl$_4$Ge$_2$ and TbAuAl$_4$Ge$_2$ crystals were cleaved in-situ, exposing the (001) crystalline plane. Typical experimental resolutions in electron energy and angle of emission were respectively 15~meV and 0.2$^{\circ}$, in all experimental setups. ARPES measurements were performed at temperatures not higher than 20~K and pressure below 5$\times$10$^{-11}$~mbar.

\begin{figure*}[t]
    \centering
    \includegraphics[width=0.98\textwidth]{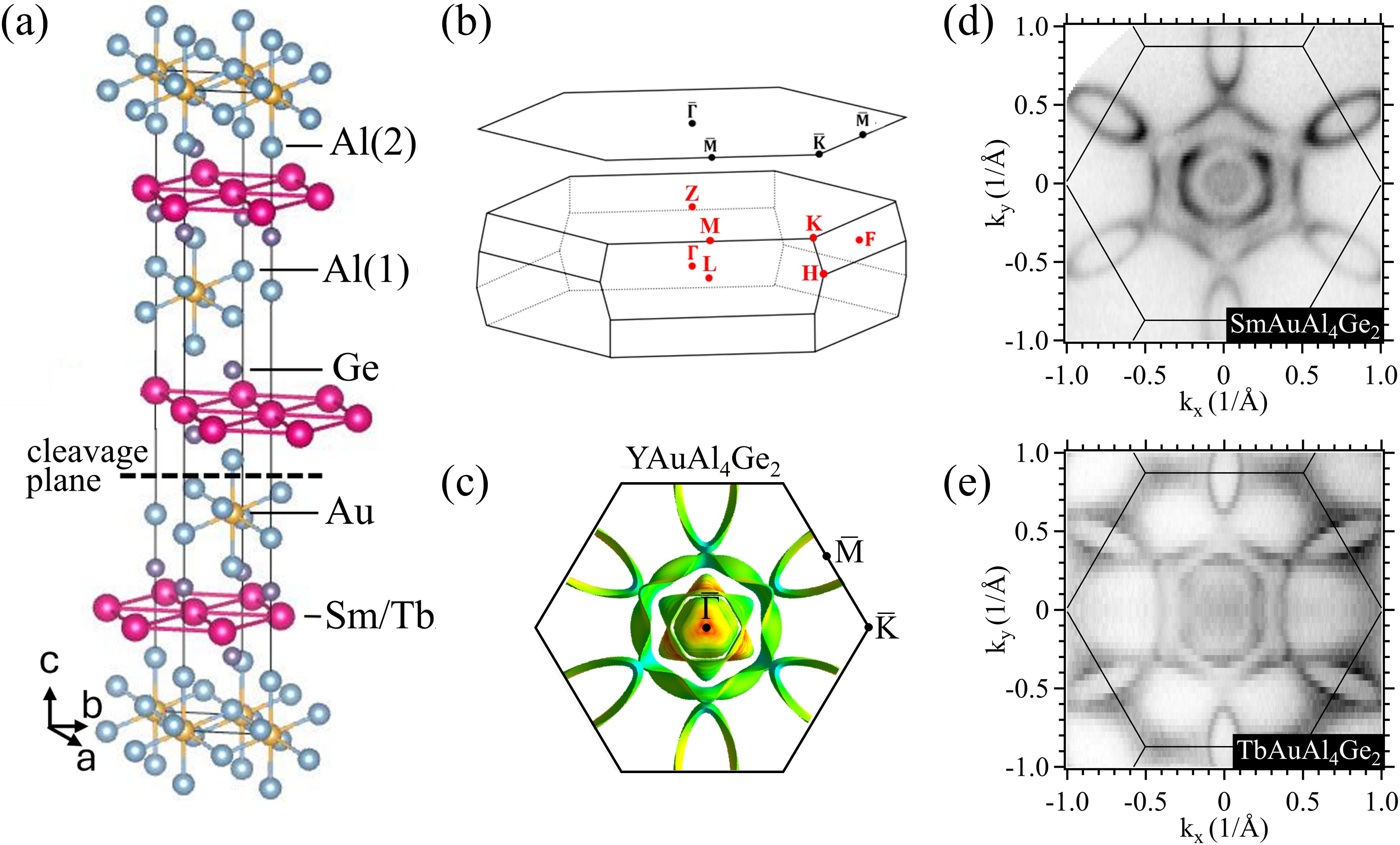}
    \caption{(a) Crystal structure of the \textit{Ln}AuAl$_4$Ge$_2$ compounds, where \textit{Ln} = Sm or Tb. There are two inequivalent sites of Al atoms, termed Al(1) and Al(2). (b) 3D Brillouin zone of \textit{Ln}AuAl$_4$Ge$_2$, its projection on the (001) plane and the corresponding high symmetry points. The plane of measurement is perpendicular to the $c$-axis. (c) Calculated Fermi surface of YAuAl$_4$Ge$_2$ obtained by density functional theory in the bulk configuration. (d) ARPES in-plane Fermi surface map (E = E$_F$) of SmAuAl$_4$Ge$_2$ reproducing the central sixfold structure and the elliptical features of panel (c). (e) ARPES in-plane Fermi surface map (E = E$_F$) of TbAuAl$_4$Ge$_2$, showing additional spectral weight along the borders of the surface Brillouin zone. The Brillouin zone boundaries are marked by faint black lines in all ARPES images. All measurements were carried out with a photon energy of 112~eV and at a temperature of 16~K. We present combined result of data acquired with photons of linear-horizontal (LH) and linear-vertical (LV) polarization.}
    \label{FIG1}
\end{figure*}

\subsection{Theoretical and Computational Approach:} \label{theory}
We performed density functional theory calculations using the generalized gradient approximation of Perdew, Burke and Ernzerhof (PBE-GGA) \cite{PhysRevLett.77.3865}, with the general potential linearized augmented planewave (LAPW) method \cite{Singh2005-wk} as implemented in the WIEN2k code \cite{10.1063/1.5143061}. We used the standard LAPW augmentation with local orbitals for semi-core states. The LAPW sphere radii were 2.55~bohr for all atoms, except Al for which 2.15~bohr was used. A well converged basis set cutoff, K$_{max}$, determined by the criterion R$_{min}$K$_{max}$=8.0, where R$_{min}$ is the radius of the smallest sphere, i.e. that of Al, was used. We used dense samplings of the Brillouin zones, and convergence with respect to the zone sampling was tested. Spin orbit coupling (SOC) was included in all calculations. We additionally used a Hubbard U=7 eV for the Tb \textit{f}-electrons. This was applied using the PBE+U method \cite{PhysRevB.57.1505}, with the fully localized limit (SIC) double counting. Calculations were done based on the experimental lattice parameters of rhombohedral TbAuAl$_4$Ge$_2$, $a$ = 4.2085~\AA, $c$ = 30.854~\AA. The internal atomic positions were relaxed using total energy minimization. This approach with a ferromagnetic Tb ordering yielded a spin moment of 6.11~$\mu_B$ per formula unit and a Tb orbital moment of 1.05~$\mu_B$ consistent with expectations for the Tb \textit{f}$^8$ shell. The orbital and spin moments are aligned parallel consistent with Hund’s rule for a more than half-filled \textit{f} shell.

We began by assessing the role of the Tb \textit{f}-electrons to the low energy (near E$_F$) electronic structure. We compared the electronic structure of TbAuAl$_4$Ge$_2$ with that of the Y analogue, YAuAl$_4$Ge$_2$ with the same crystal structure as shown in Fig. \ref{FIG1}. The occupied Tb \textit{f}-states are at high binding energy -10~eV to -5~eV relative to E$_F$, and do not affect the electronic structure in the range of our ARPES measurements. The lowest unoccupied Tb \textit{f} states are at $\sim$2~eV above E$_F$. The Au \textit{d} shell is fully occupied in both compounds with the Au \textit{d} states appearing in the range of approximately -8~eV to -6~eV. In view of this we calculated the electronic structure of slabs of YAuAl$_4$Ge$_2$, which computationally more tractable and provides for a simpler analysis due to the absence of \textit{f}-electron magnetism.

We constructed slab models with different surface terminations to compare with the ARPES results. These were constructed by going from the primitive one formula unit cell, to the conventional hexagonal cell and doubling along the $c$-axis to form slabs of six layers of YAuAl$_4$Ge$_2$ with added vacuum layers. The surface termination was changed by adding layers to the slabs. The in-plane lattice parameter was kept fixed at the experimental value, and the atomic positions were relaxed by total energy minimization. All slabs were constructed with hexagonal symmetry, and inversion symmetry about the center of the slab.

\section{Results and Discussion}
SmAuAl$_4$Ge$_2$ and TbAuAl$_4$Ge$_2$ crystallize in a trigonal structure (space group no. 166) \cite{PhysRevB.109.014436} formed by Sm/Tb staggered layers of two-dimensional triangular lattices stacked along the c-axis, the triangular atomic arrangement satisfies the geometric conditions for magnetic frustration. The lattice parameters for Sm(Tb)AuAl$_4$Ge$_2$ are found to be $a = b =$ 4.217(4.2085)~\r{A}, $c =$ 31.145(30.854)~\r{A} \cite{PhysRevB.109.014436, feng_magnetic_2022}. In Figs. \ref{FIG1}(a) and \ref{FIG1}(b) we, respectively, show the crystal unit cell and the first Brillouin zone of the \textit{Ln}AuAl$_4$Ge$_2$ compounds.

\begin{figure*}[t]
    \centering
    \includegraphics[width=0.98\textwidth]{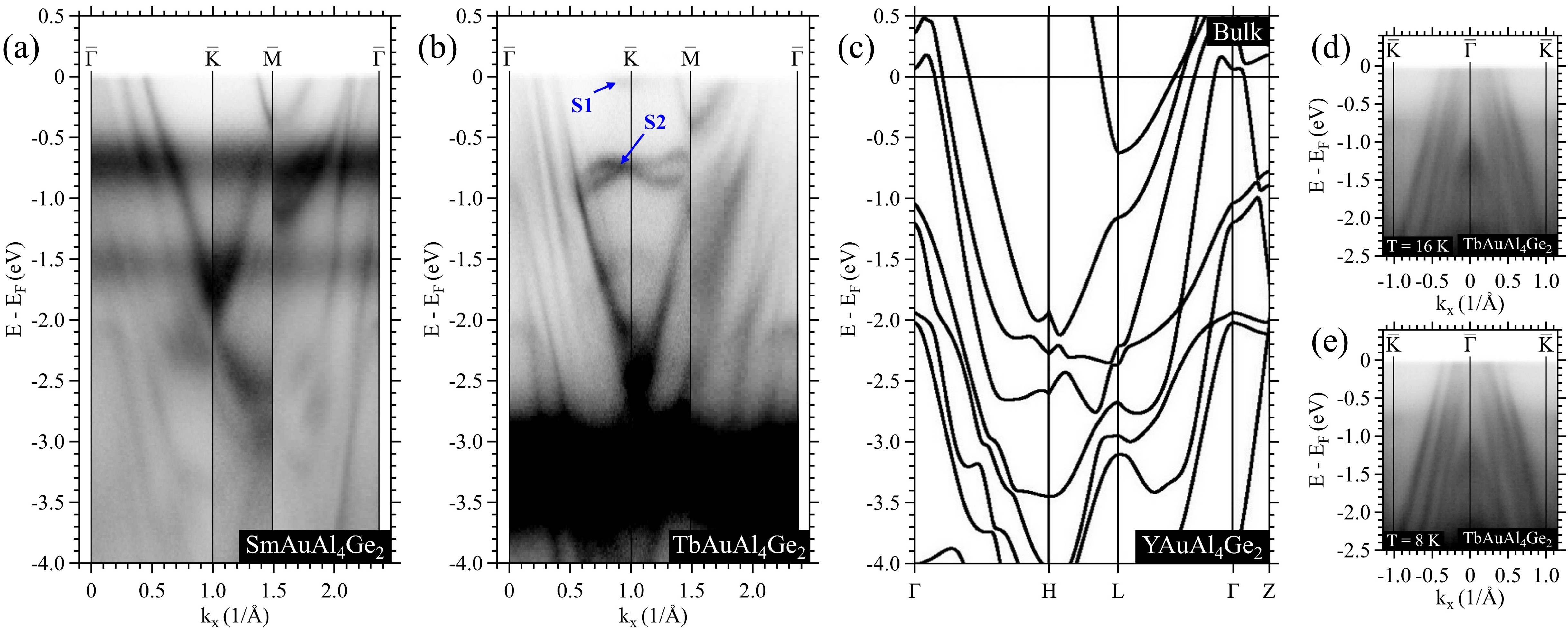}
    \caption{(a) ARPES energy-momentum map of SmAuAl$_4$Ge$_2$ along $\bar{\Gamma}$-$\bar{K}$-$\bar{M}$-$\bar{\Gamma}$, showing the dispersion of the electronic states in the near-E$_F$ region. The two non-dispersive states are due to Sm 4\textit{f} electrons. (b) ARPES energy-momentum maps of TbAuAl$_4$Ge$_2$ along $\bar{\Gamma}$-$\bar{K}$-$\bar{M}$-$\bar{\Gamma}$, showing the dispersion of the electronic states in the near-E$_F$ region. The intense non-dispersive feature at around -3.3~eV arises from Tb 4\textit{f} electrons. Both maps were measured at a temperature of 16~K. (c) Calculated band structure of YAuAl$_4$Ge$_2$ computed by DFT in the bulk configuration along the reciprocal path $\Gamma$-$H$-$L$-$\Gamma$-$Z$. (d),(e) ARPES energy-momentum map of TbAuAl$_4$Ge$_2$ along $\bar{K}$-$\bar{\Gamma}$-$\bar{K}$ acquired at respectively, T = 16~K (above the transition temperature) and T = 8~K (below the transition temperature). The two maps show that, within our experimental resolution, 
    there are no changes in the electronic structure across the transition. All measurements were carried out with a photon energy of 112~eV using linear-horizontal (LH) polarization.}
    \label{FIG2}
\end{figure*}

\subsection{The electronic structure of Sm(Tb)AuAl\texorpdfstring{$_4$}{4}Ge\texorpdfstring{$_2$}{2}} \label{bandstructure}

The calculated Fermi surface for the family of these compounds is shown in Fig. \ref{FIG1}(c) from the viewpoint of the (001) plane. The contours follow the symmetry of the Brillouin zone. As explained in section \ref{theory}, the lanthanide atoms have been replaced by Y atoms for the calculations. The calculated densities of states of the Y compounds show no differences with the lanthanide analogues, while the \textit{f}-derived bands do not affect the principal features of the electronic structure near the Fermi level (see supplementary info SI1). Therefore, by replacing the lanthanide atoms with Y, one can have access to the complete electronic structure below E$_F$ without the latter being masked by the 4\textit{f} states of the lanthanides. 

The experimental Fermi surfaces of SmAuAl$_4$Ge$_2$ and TbAuAl$_4$Ge$_2$ are shown in Figs. \ref{FIG1}(d) and \ref{FIG1}(e). We note that these are the in-plane Fermi surfaces measured at an out-of-plane k-value that is determined by the photon energy (112~eV), and hence correspond to a 2D cut of the FS at around $\Gamma$. The features shown in Fig. \ref{FIG1}(c) are well reproduced. First of all, there are three experimental contours around $\bar{\Gamma}$ with the sixfold symmetry becoming more apparent for the outer one. Further away from $\bar{\Gamma}$, this outer hexagram evolves as six elliptical patterns that cross the boundaries of the surface Brillouin zone, with each one centred at $\bar{M}$. The abundance of Fermi surface contours is consistent with the transport properties of the two compounds, which remain good metals at all temperatures \cite{feng_magnetic_2022,PhysRevB.109.014436}.

The experimental energy-momentum dispersion of the electronic states (i.e. the electronic structure) is shown in Figs. \ref{FIG2}(a) and \ref{FIG2}(b). These maps follow the experimental dispersion along the $\bar{\Gamma}$-$\bar{K}$-$\bar{M}$-$\bar{\Gamma}$ path of the surface Brillouin zone. Along the $\bar{\Gamma}$-$\bar{K}$ direction, one can readily observe three hole-like (concavity down) metallic states centered around $\bar{\Gamma}$. These are the states that form the sixfold contours previously shown in Figs. \ref{FIG1}(d) and \ref{FIG1}(e) with the dispersion of the outer one displaying a band minimum at $\bar{K}$ at an energy of 1.7~eV (2.3~eV) for the Sm (Tb) compound. Moreover, there is an electron-like (concavity up) metallic state centered at the $\bar{M}$ point that forms the elliptical pockets of Figs. \ref{FIG1}(d) and \ref{FIG1}(e). There are in total one $\bar{\Gamma}$ and three equivalent $\bar{M}$ points per Surface Brillouin zone (SBZ). Therefore, the 2D Fermi surface consists of three hole pockets (around $\bar{\Gamma}$) and three electron pockets (around $\bar{M}$). Although, the states straddling the Fermi level are the most important ones since they determine the physical properties of the compounds, we can experimentally probe at least three other electronic states with higher binding energies centered at $\bar{\Gamma}$. Two hole-like states with energy maxima of 1.0~eV and a third hole-like state with energy maxima of 2.1~eV, more clearly resolved in TbAuAl$_4$Ge$_2$. Finally, there is another electron-like state with a minimum of around 1.2~eV centered at $\bar{M}$. All these features are largely reproduced by the bulk DFT calculations shown in Fig. \ref{FIG2}(c). First of all, one can easily observe the three hole-like metallic states along the $\Gamma$-$H$ path. Moreover, there are two sets of hole-like states at $\Gamma$ and $Z$ found at higher binding energies and two electron-like states centered at $L$. Given that $\Gamma$, $Z$ and $L$ are projected into $\bar{\Gamma}$, $\bar{\Gamma}$ and $\bar{M}$, respectively [see Fig. \ref{FIG1}(b)], the theoretical binding energies of the corresponding band maxima and minima are in excellent agreement with the experimental values.

\begin{figure*}[t]
    \centering
    \includegraphics[width=0.98\textwidth]{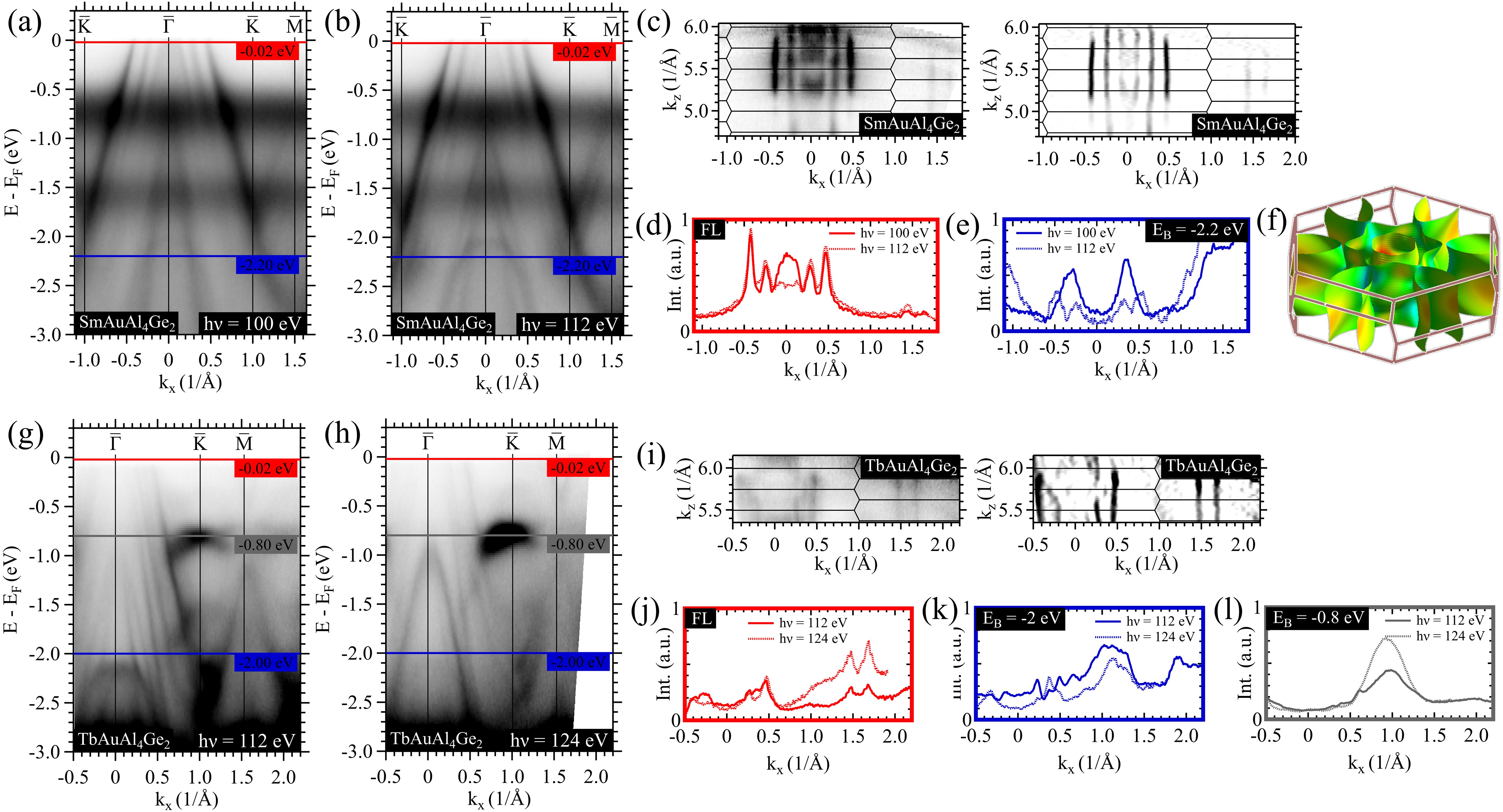}
    \caption{(a),(b) ARPES energy-momentum map of SmAuAl$_4$Ge$_2$ along $\bar{\Gamma}$-$\bar{K}$-$\bar{M}$ measured with a photon energy of 100~eV and 112~eV, respectively. (c) ARPES out-of-plane Fermi surface map (E = E$_F$) of SmAuAl$_4$Ge$_2$ measured along $\Gamma$-Z, showing the out-of-plane dispersion of the metallic states. The measurements were carried out using photon energies from 70 to 130~eV. Left: raw data. Right: data processed using the curvature method \cite{Zhang2011APM} to enhance the intensity of weak spectral features. (d) Momentum distribution curves of SmAuAl$_4$Ge$_2$ near the Fermi level, obtained from the energy-momentum maps in (a) and (b), evidencing the dispersive character of the inner metallic state and the non-dispersive character of the outer metallic states. (e) Momentum dispersive curves of SmAuAl$_4$Ge$_2$ at E$_B\approx-2.2$~eV, obtained from the energy-momentum maps in (a) and (b), evidencing the dispersive character of the valence band. The energy lines were the MDC curves were obtained are shown in images (a) and (b). (f) Theoretical Fermi surface (E = 0) of YAuAl$_4$Ge$_2$, showing the strong 2D character of the metallic states. (g) ARPES energy-momentum map of TbAuAl$_4$Ge$_2$ along $\bar{\Gamma}$-$\bar{K}$-$\bar{M}$ measured with a photon energy of 112~eV and 124~eV, respectively. (i) ARPES out-of-plane Fermi surface map (E = E$_F$) of TbAuAl$_4$Ge$_2$ measured along $\Gamma$-Z, showing the out-of-plane dispersion of the metallic states. The measurements were carried out using photon energies from 100 to 150~eV. Left: raw data. Right: data processed using the curvature method. (j) Momentum distribution curves of TbAuAl$_4$Ge$_2$ near the Fermi level, obtained from the energy-momentum maps in (g) and (h). (k) Momentum distribution curves of TbAuAl$_4$Ge$_2$ at E$_B\approx-2$~eV, obtained from the energy-momentum maps in (g) and (h), evidencing the dispersive character of the valence band. The energy lines were the MDC curves were obtained are shown in images (g) and (h). (l) Momentum dispersive curves of TbAuAl$_4$Ge$_2$ at E$_B\approx-0.8$~eV, obtained from the energy-momentum maps in (g) and (h), evidencing the non-dispersive character of the state S2. All measurements were carried out using linear-horizontal polarization and at a temperature of 16~K}
    \label{FIG3}
\end{figure*}

An obvious difference between the two dispersion maps is the energy value of the non-dispersive states that represent the photoemission peaks of the 4\textit{f} states of the lanthanide ion. These are located at 0.7~eV and 1.5~eV in the case of Sm (4\textit{f} 7/2 and 4\textit{f} 5/2 doublet), and at around 3.3~eV in the case of Tb. These 4\textit{f}-derived features are not observed in the theoretical band structure of Fig. \ref{FIG2}(c) because the lanthanide atoms have been replaced by Y as explained previously. Besides the 4\textit{f} peaks, there are some other notable differences in the electronic dispersion of the two compounds. In the case of TbAuAl$_4$Ge$_2$ there are features centered at the $\bar{K}$ point of the SBZ that have no counterpart in SmAuAl$_4$Ge$_2$, and most importantly, they are not reproduced in bulk calculations. These are a weakly dispersing state at an approximate energy of 0.75~eV that is made from two crossing branches and a weak diffuse state right at the Fermi level. The latter gives rise to the diffuse intensity ring that follows the borders of the SBZ in the FS of TbAuAl$_4$Ge$_2$ [see Fig. \ref{FIG1}(e)]. Before discussing the origin of these states in section \ref{terminations}, we will focus on the temperature dependence and the dimensionality of the electronic structure. 

The motivation to examine the temperature-dependent electronic structure arises from the fact that both compounds exhibit a double magnetic transition at low temperatures: SmAuAl$_4$Ge$_2$ at 13.2~K and 7.4~K, and TbAuAl$_4$Ge$_2$ at 13.9~K and 9.8~K. These transitions have been well documented in recent literature by means of magneto-transport and lead to complex magnetic ordering \cite{feng_magnetic_2022,PhysRevB.109.014436}. A new periodicity due to magnetic ordering will be theoretically reflected into the periodicity of the Brillouin zone and hence on the electronic structure, for example by introducing band replicas following that periodicity or new band gaps at the borders of the modified Brillouin zone. In Figs. \ref{FIG2}(d) and \ref{FIG2}(e), we present the energy-momentum dispersion of the electronic states of TbAuAl$_4$Ge$_2$ near the Fermi level at two different temperatures: 16~K and 8~K, respectively. We did not observe any temperature-dependent differences, either in the bulk electronic dispersion, or in the Fermi surface including its fine details. This is no surprising as, in an ARPES experiment, the intensity of any replicated bands is usually very low with respect to the parent bands, while the size of an induced energy gap at such temperatures would be beyond our experimental energy resolution. An additional complication is of course the complex magnetic ordering at low temperatures that does not allow for a simple definition of a new periodicity. In the following, we will turn to the dimensionality of the electronic structure.

We have previously discussed the presence of six Fermi pockets per surface Brillouin zone in both materials. Three of them form the sixfold structure centered at $\bar{\Gamma}$ and the remaining correspond to the elliptical contours around the six equivalent $\bar{M}$ points. In order to analyze the dimensionality of the metallic states forming the Fermi surface contours, i.e. whether they are 2D or 3D states, energy-momentum maps along $\bar{\Gamma}$-$\bar{K}$-$\bar{M}$ were obtained by varying the photon energy used for each measurement. Two exemplary energy-momentum maps for each compound are shown in panels (a), (b), (g) and (h) of Fig. \ref{FIG3}. In the case of SmAuAl$_4$Ge$_2$, a quick comparison of Figs. \ref{FIG3}(a) and \ref{FIG3}(b) shows no major differences in the dispersion of the metallic states, thereby suggesting the absence of out-of-plane dispersion. Nevertheless, before drawing any final conclusion, one must follow the out-of-plane dispersion along -at least- one complete Brillouin zone. Therefore, using energy-momentum maps acquired with photon energies from 70~eV to 130~eV, the out-of-plane Fermi surface has been constructed and the result is shown in Fig. \ref{FIG3}(c). The innermost state centered at $\bar{\Gamma}$ shows signs of an out-of-plane dispersion and a potentially closed contour. However, large intensity variations hinder its observation along a large k$_z$-range. This is the only metallic state with potential signs of an out-of-plane dispersion, as all other states centered at $\bar{\Gamma}$ [left part of Fig. \ref{FIG3}(c)] or $\bar{M}$ [right part of Fig. \ref{FIG3}(c)] yield open FS contours and they are therefore 2D. To cast away any further doubt on the out-of-plane dispersion of the metallic states, momentum distribution curves (MDCs) were extracted from the maps of Figs. \ref{FIG3}(a) and \ref{FIG3}(b). Two curves were plotted for each map, one near the Fermi level and the other at an approximate binding energy of 2.2~eV, as indicated by the horizontal lines in both figures. The MDCs are shown in Figs. \ref{FIG3}(d) and \ref{FIG3}(e). Near the Fermi level, the peaks corresponding to the outer states that are centred at $\bar{\Gamma}$ overlap with their counterparts measured at a different photon energy. This is again a clear signature of no dispersion in the out-of-plane direction, with the sole exception of the innermost state that shows spectral weight only in the 100~eV measurement. The lack of out-of plane dispersion is also confirmed for the state centred at $\bar{M}$ (peak at 1.5~\AA$^{-1}$). On the other hand, the MDCs extracted at 2.2~eV show substantial changes in the peaks’ positions, thereby, revealing the 3D character of the electronic states lying at lower energies.

The same procedure has been followed to study the dimensionality of the metallic states in TbAuAl$_4$Ge$_2$. The experimental out-of-plane Fermi surface has been constructed with energy-momentum maps acquired with photon energies from 100~eV to 150~eV [Fig. \ref{FIG3}(i)] and two exemplary maps are shown in Figs. \ref{FIG3}(g) and \ref{FIG3}(h). As in the samarium analogue, the energy-momentum maps of TbAuAl$_4$Ge$_2$ show identical dispersion near the Fermi level, indicating the absence of out-of-plane dispersion for the experimentally observed metallic states. The innermost metallic state which showed potential signs of an out-of-plane dispersion in SmAuAl$_4$Ge$_2$ cannot be clearly observed in the case of TbAuAl$_4$Ge$_2$. On the other hand, and again in line with our observations on the Sm compound, a mere inspection of Figs. \ref{FIG3}(g) and \ref{FIG3}(h) shows clear changes in the dispersion of the electronic states lying in higher binding energies. In order to verify that the metallic states in TbAuAl$_4$Ge$_2$ have a 2D character, while states lying deeper in energy are mainly 3D, MDC have been extracted from the energy-momentum maps of Figs. \ref{FIG3}(g) and \ref{FIG3}(h). The results are shown in Figs. \ref{FIG3}(j)-(l). The intensity peaks corresponding to the metallic states centered at $\bar{\Gamma}$ and $\bar{M}$ do not shift to different k-values when h$\nu$ changes, signalling the absence of out-of-plane dispersion [Fig. \ref{FIG3}(j)]. On the other hand, clear signs of dispersion are observed on the photoemission peaks extracted at 2.0~eV [Fig. \ref{FIG3}(k)]. Finally, Fig. \ref{FIG3}(l) tracks the MDCs corresponding to the highly intense feature centered at $\bar{K}$ at an approximate binding energy of 0.75~eV. This feature was briefly mentioned in the frame of Fig. \ref{FIG2} and, most importantly, it is not reproduced by bulk calculations. The corresponding MDCs show mere intensity differences but no signs of dispersion as a function of the photon energy. Therefore, this feature must also correspond to a state of lower dimensionality and it will be further discussed in the following section.

\begin{figure*}[t]
    \centering
    \includegraphics[width=0.98\textwidth]{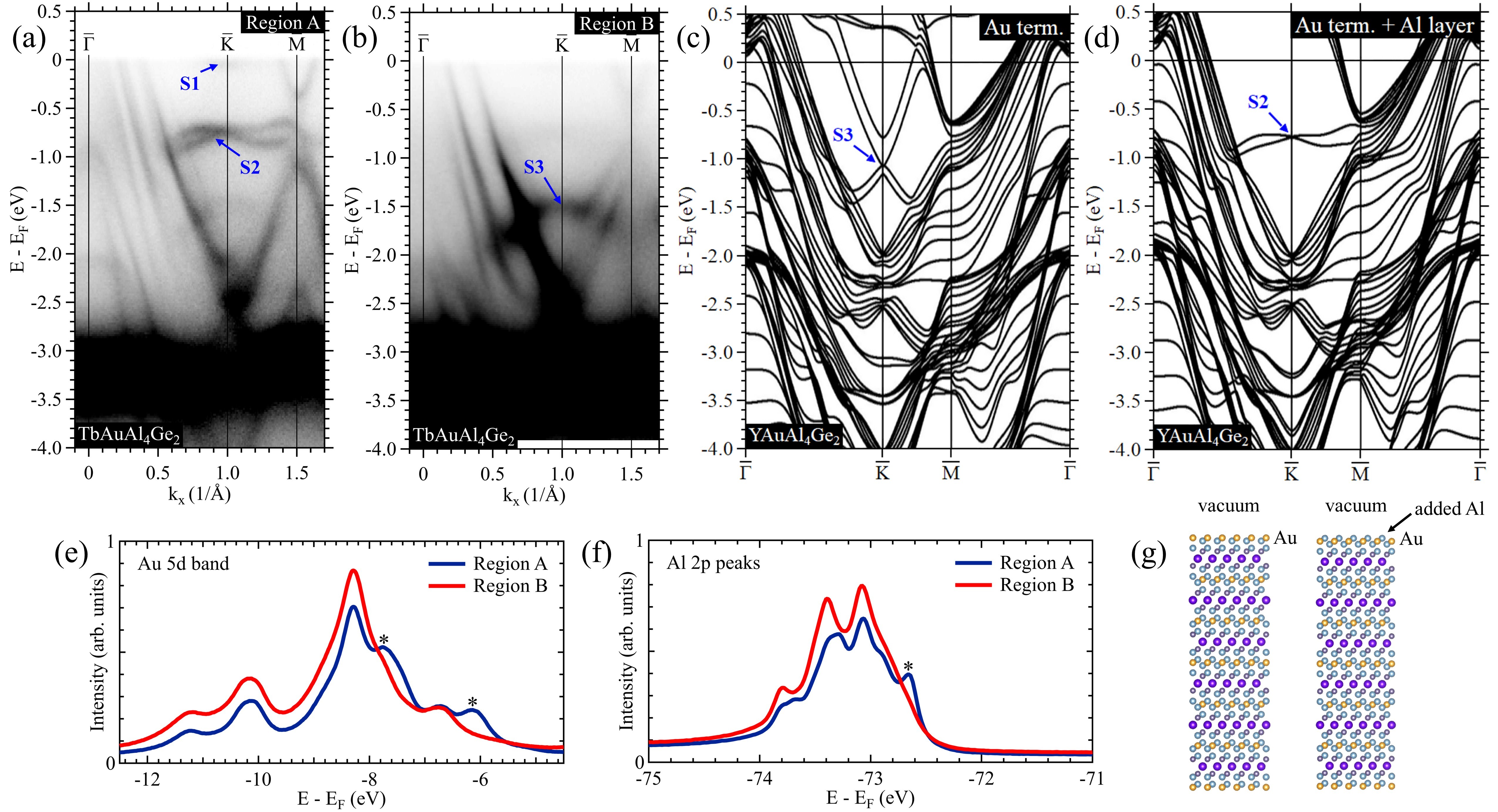}
    \caption{(a),(b) ARPES energy-momentum maps of TbAuAl$_4$Ge$_2$, respectively in region A and region B, along $\bar{\Gamma}$-$\bar{K}$-$\bar{M}$ showing the near-E$_F$ electronic structure. (c),(d) Band structure of YAuAl$_4$Ge$_2$ along $\bar{\Gamma}$-$\bar{K}$-$\bar{M}$-$\bar{\Gamma}$, computed by DFT, respectively in the slab configuration terminating in the Au atomic layer and in the Au atomic layer with an Al monolayer on top. The S2 surface feature is seen at around E=-0.75~eV. (e) Angle-integrated intensity map of the Au 5d state in regions A and B in TbAuAl$_4$Ge$_2$. The extra peaks observed in region A are marked with an asterisk. (f) Angle-integrated intensity map of the Al 2p core level peak in regions A and B of TbAuAl$_4$Ge$_2$. The extra peak observed in region A is marked with an asterisk. (g) Graphical representation of the slab configuration used in (c) (left) and (d) (right). All measurements were carried out at 16~K with a photon energy of 112~eV using linear-horizontal polarization}
    \label{FIG4}
\end{figure*}

All experimental evidence therefore points towards a highly two-dimensional Fermi surface formed by the bulk electronic states around the $\bar{\Gamma}$ and $\bar{M}$ high-symmetry points of the SBZ. Electronic states lying deeper in energy are rather 3D, with the notable exception of the intense feature centered at $\bar{K}$ and not reproduced by bulk calculations. The lack of observable out-plane dispersion for the bulk states is in line with the layered structure of the compounds and with previous ARPES results on GdAuAl$_4$Ge$_2$ \cite{zhang_multiple_2023}. We note that bulk DFT calculations shown in Fig. \ref{FIG2}(c), reveal very small dispersion along the $\Gamma$-$Z$ direction of the 3D Brillouin zone within the first 2~eV below the Fermi level, with the single exception of one band that crosses E$_F$ along $\Gamma$-$Z$ and may correspond to the innermost experimental contour around $\bar{\Gamma}$. The theoretical Fermi surface of YAuAl$_4$Ge$_2$ shown in Fig. \ref{FIG3}(f) is very clear: the two outermost features around $\bar{\Gamma}$, as well as the elliptical features around $\bar{M}$ are both shown to form open Fermi surface contours and they are therefore 2D in excellent agreement with our experimental findings.

Having discussed in detail the bulk electronic structure of SmAuAl$_4$Ge$_2$ and TbAuAl$_4$Ge$_2$, we will now focus on features of their electronic structure that are not reproduced by bulk calculations. Through the combined results of ARPES, angle-integrated photoemission and slab calculations, we will reveal the termination layer of Sm(Tb)AuAl$_4$Ge$_2$ single crystals and its effect on the electronic structure.

\subsection{Revealing the surface termination layer} \label{terminations}

Figs. \ref{FIG4}(a) and \ref{FIG4}(b) show two typical energy-momentum maps of TbAuAl$_4$Ge$_2$ along $\bar{\Gamma}$-$\bar{K}$-$\bar{M}$ measured at different sample locations. Although the main experimental features are identical, it is clear that Fig. \ref{FIG4}(a) displays a ``richer'' electronic structure with some states having no counterpart in Fig. \ref{FIG4}(b). In the following we will refer to these results as ``region A'' [Fig. \ref{FIG4}(a)] and ``region B'' [Fig. \ref{FIG4}(b)]. First of all, one can readily observe in Fig. \ref{FIG4}(a) two secondary features at the $\bar{K}$ point of the SBZ: a faint feature at the Fermi level that gives rise to the diffuse intensity ring in the experimental FS of Fig. \ref{FIG1}(e) and an intense x-like feature at around 0.75~eV. In the following, these features will be, respectively, termed ``S1'' and ``S2''. Moreover, there are differences in the relative intensity of the hole-like states centered at $\bar{\Gamma}$ with most of them being more clearly resolved in Region A. These differences in the electronic structure between the two regions reveal that the termination layer of the cleaved TbAuAl$_4$Ge$_2$ is not homogeneous. In the following, we will reveal the dominant termination layer through the comparison of these experimental results to DFT slab calculations.

In a related study on GdAuAl$_4$Ge$_2$, Zhang and coworkers have probed by means of ARPES several features centred at the $\bar{K}$ point of SBZ that are not predicted by bulk calculations \cite{zhang_multiple_2023}. Through a comparison to surface-sensitive calculations, the authors concluded that the termination layer of GdAuAl$_4$Ge$_2$ cleaved single crystals is the Au layer. Since the lanthanide atom is not expected to affect the crystal structure, our first choice for the cleavage plane -and hence the termination layer- of the TbAuAl$_4$Ge$_2$ compounds was naturally the Au layer. In order to check this hypothesis, we compared DFT slab calculations with the experimental results acquired on the surface region that presents the most obvious electronic states that are not related to the bulk (i.e. region A).

Fig. \ref{FIG4}(c) shows the electronic structure of a finite YAuAl$_4$Ge$_2$ slab with Au termination as shown in the inset. As in the case of the bulk calculation, the replacement of the lanthanide atoms with Y does not affect the main features of the electronic structure near the Fermi level, and permits having access to the near-E$_F$ energy-momentum dispersion without the latter being partially hidden behind the intense 4\textit{f} states of the lanthanides. The main experimental features are well reproduced by three principal sets of hole-like bands at $\bar{\Gamma}$: at around E$_F$, at 1~eV and at 2~eV. The faint lines between these three main sets of bulk bands are artifacts due the finite thickness of the slab. To be more specific, the states corresponding to the bulk continuum are seen as bands with almost identical dispersions and small energy shifts within the bulk continuum. Hence, energy states, that have no counterpart in the bulk continuum correspond to non-bulk bands, i.e. surface states. An example, are the features marked by arrows within the bulk gap at the $\bar{K}$ point. However, the features predicted by the Au-terminated slab calculation [Fig. \ref{FIG4}(c)] do not match the experimental dispersion of S1 and S2. One may therefore conclude that the surface termination layer in Region A cannot be simply attributed to a Au layer.

In order to acquire further experimental input, we compared the lineshape of the Au 5\textit{d} and Al 2\textit{p} photoemission peaks for the two regions. The comparison of the Au 5\textit{d} peaks is shown in Fig. \ref{FIG4}(e): interestingly, the lineshape of region A exhibits two secondary peaks that have no match in the lineshape of region B. This is a proof that Au atoms are found in a larger variety of chemical environments in region A than in region B, and thus a good indication for Au surface atoms in the former. As explained previously, a simple Au termination is, however, in very poor agreement with the predictions of our slab calculations. We therefore compared the Al 2\textit{p} lineshape in the two regions and we, likewise, observed that Al atoms are found in a larger variety of chemical environments in region A. The secondary peak of Al 2\textit{p} is marked by an asterisk in Fig. \ref{FIG4}(f). As a matter of fact, Figs. \ref{FIG4}(e) and \ref{FIG4}(f) suggest that both Au and Al may reside on the surface. 

In Fig. \ref{FIG4}(d) we present the results of another DFT calculation where a monolayer of Al is added on the Au-terminated slab. This calculation simulates a cleavage plane for the Tb(Sm)AuAl$_4$Ge$_2$ just above the Al(1) atoms of Fig. \ref{FIG1}(a). In this way, both Au and Al atoms are exposed on the surface termination layer. The theoretical energy-momentum dispersion is now in very good agreement with the experimental data on region A. The S2 feature at $\bar{K}$ is perfectly captured and it therefore corresponds to a surface state. The orbital projection of our slab calculations reveals that S2 is related to the out-of-plane \textit{p} orbitals of Au (see Fig. SF3 in supplementary info SI2). Moreover, the three sets of bulk bands around $\bar{\Gamma}$ and the electron pocket at $\bar{M}$ remain in very good agreement with the experimental findings. The S1 state is the only experimental feature that is not reproduced. A similar feature had been reported on GdAuAl$_4$Ge$_2$ and it was then attributed to a purely Au-terminated surface \cite{zhang_multiple_2023}. However, as can be seen in Fig. \ref{FIG4}(c), we cannot draw such a conclusion from the Au-terminated slab. We cannot exclude the possibility that S1 would be theoretically predicted in a slab calculated with Tb instead of the non-magnetic Y, but we note again that the inclusion of Tb \textit{f} moments has no impact on the electronic density of states near the Fermi level (see Fig. SF1 in supplementary info SI1). Despite this discrepancy, the attribution of the cleavage surface to a plane terminated by the Al(1) atoms stays in very good agreement with the DFT calculations and the indications from angle-integrated photoemission [Figs. \ref{FIG4}(e) and \ref{FIG4}(f)]. Region A therefore corresponds to a well-ordered surface region of a TbAuAl$_4$Ge$_2$ single crystal that has been cleaved along the plane between the Al(1) and Al(2) atoms [marked in Fig. \ref{FIG1}(a)].

\begin{figure*}[t]
    \centering
    \includegraphics[width=0.98\textwidth]{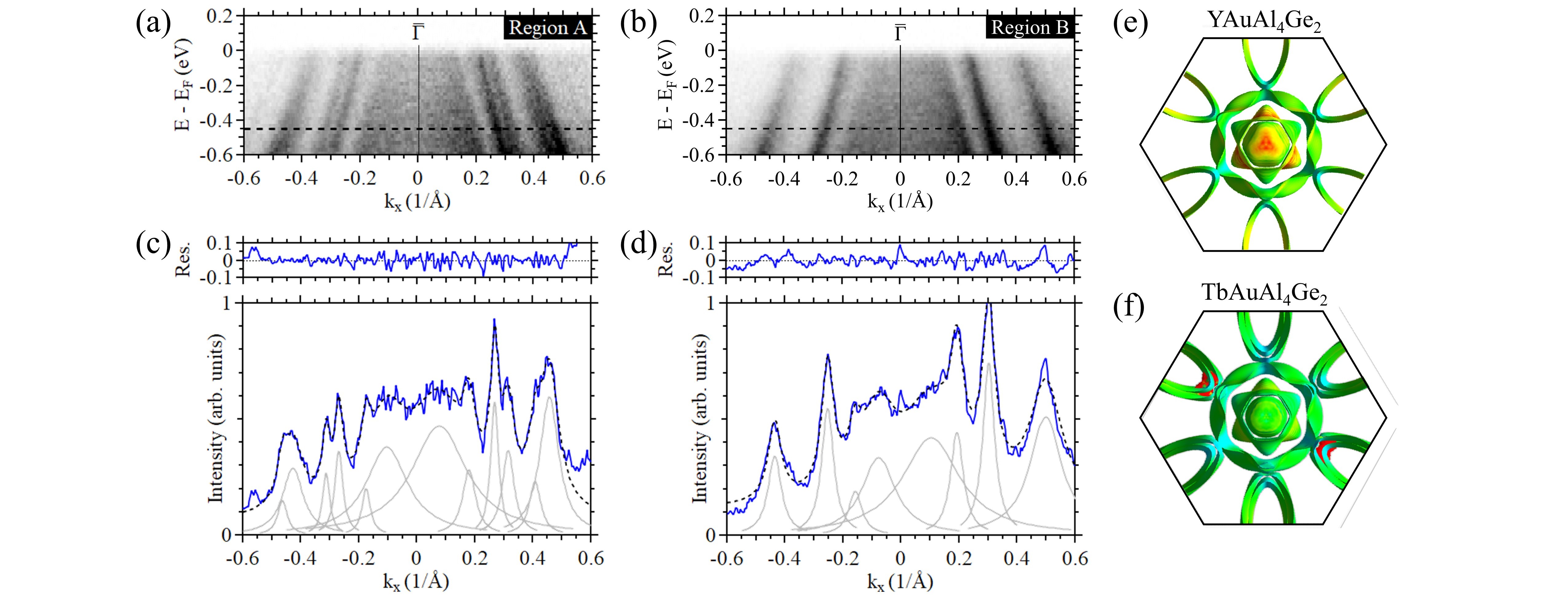}
    \caption{(a), (b) Energy-momentum dispersions along the $\bar{K}$-$\bar{\Gamma}$-$\bar{K}$ high symmetry direction for region A and B respectively. (c), (d) Momentum distribution curves (MDC) of (a) and (b) respectively where the dashed line at E - E$_F$ = 450~meV indicates the position of these MDCs. The integration range is set to 40 meV and we used lorentzians peaks and a constant background to perform the fits. The residual signal of the fit is shown in the top most graphs of (c) and (d). (e), (f) Theoretical constant energy (E = E$_F$) maps for TbAuAl$_4$Ge$_2$ and  for YAuAl$_4$Ge$_2$ with exchange and spin-orbit coupling taken into account for the Tb compound.}
    \label{FIG5}
\end{figure*}

Although the dominant surface structure in well-ordered regions is Al(1)-terminated, this seems to be far from a natural cleavage plane. Next to the well-ordered region A, we have observed areas where clear surface features are lacking and even bulk features are less well-resolved (i.e. so-called region B). Region B may be therefore attributed to a rougher surface area -repeatedly- observed in the same cleavage surface. Interestingly, within region B, there are certain areas where a different feature appears in the bulk gap region of the $\bar{K}$ high-symmetry point. This feature lies at an approximate binding energy of 1.5~eV at $\bar{K}$ and it is pointed by an arrow in Fig. \ref{FIG4}(b). The corresponding state, termed S3, is only barely visible on region A. The energy range of S3 fairly matches the lower part of the calculated surface state pointed by an arrow in the slab calculation of Fig. \ref{FIG4}(c).

The theoretical surface state originates from the in-plane \textit{p} orbitals of Au (see Fig. SF3 in supplementary info SI2), but, experimentally, neither its upper part, nor its detailed dispersion can be probed. The only experimental signature is a blurry feature at the energy range where the dispersion is expected to be weaker (i.e. the lower part). Although the assignment of S3 to the surface state of a Au-terminated surface remains uncertain, extinction/smearing of the experimental energy-momentum dispersion of surface-derived features in rough areas is a reasonable possibility. We note that the calculated surface state is very localised in the near-surface region (see Fig. SF2 in supplementary info SI2) and therefore prone to surface disorder. The rough region B might therefore correspond to areas where a part of the surface Al(1) atoms is missing and the surface is terminated by a Au layer. The random absence of Al(1) atoms makes some areas of our cleavage surface rougher than the well-ordered Al(1)-terminated regions. In short, we attribute those rougher areas to region B and the well-ordered areas to region A. Our work therefore suggests that although the most favorable cleavage plane is above the Al(1) atoms, the surface of TbAuAl$_4$Ge$_2$ remains rough at scales comparable to a typical ARPES beam spot and invites future investigations with micro/nano-metric probes.

\subsection{The effect of localized  magnetic moments} \label{magneticmoments}

There is an interesting observation about the electronic structure around the Fermi level in the two regions. Figs. 5(a) and 5(b) present a zoom near the Fermi level for the data already shown in Figs. 4(a) and 4(b), respectively. In region A [panel (a)], each hole-like band around $\bar{\Gamma}$ is split into two states in the near-E$_F$ energy range, and it therefore exhibits a replica which crosses the Fermi level at a neighbouring k-value. The splitting is more obvious in the momentum distributions curves of Figs. \ref{FIG5}(c) and \ref{FIG5}(d), where each peak is effectively doubled at the Fermi level for region A. The MDC is integrated in the range [-0.43,-0.47] eV along the $\bar{K}$-$\Gamma$-$\bar{K}$ direction and we use lorentzians peaks with a constant background to fit the different peaks. 

We attribute the experimentally-observed band splitting to the effect of the exchange and spin-orbit interactions in TbAuAl$_4$Ge$_2$. Up to now, for reasons of simplicity, we have been comparing the electronic structure of Tb(Sm)AuAl$_4$Ge$_2$ to YAuAl$_4$Ge$_2$ because the 4\textit{f} electrons of the lanthanides do not hybridise with the rest of the near-E$_F$ electronic structure. The Tb \textit{f} electrons possess however spin magnetic moments that lead to an asymmetry between the majority and the minority spins in TbAuAl$_4$Ge$_2$ (see supplementary info SI1). The spins of the conduction electrons may interact with the spins of those localized  4\textit{f} magnetic moments through the exchange interaction. The combined effect of this exchange interaction and the spin-orbit interaction of the conduction electrons leads to fine energy splitting of the near-E$_F$ electronic structure. This is apparent through the comparison of the calculated Fermi surfaces of TbAuAl$_4$Ge$_2$ [Fig. \ref{FIG5}(e)] and YAuAl$_4$Ge$_2$ [Fig. \ref{FIG5}(f)]. Each Fermi sheet exhibits a clear splitting in the case of TbAuAl$_4$Ge$_2$, a splitting which is absent for the Y analogue where no \textit{f} electrons are present. This splitting is a bulk effect and should be present all over the cleavage surface. However, in rougher surface areas where the sample surface does not show a long range crystalline order (i.e. region B), the splitting is not apparent just like other details of the electronic structure (see beginning of section \ref{terminations}). TbAuAl$_4$Ge$_2$ is therefore a good example of how the \textit{f} electrons can modify the Fermi surface, and hence the conduction properties of a compound, without interacting directly with the conduction electrons other than with their spin degree of freedom.

\section{Summary and Conclusions}
The experimental electronic structure of SmAuAl$_4$Ge$_2$ and TbAuAl$_4$Ge$_2$ can be well captured by their non-magnetic analogue YAuAl$_4$Ge$_2$, in line with the expected negligible hybridisation between the \textit{f} electrons and the conduction band. It shows a quasi-two-dimensional Fermi surface formed by multiple hole and electron pockets in agreement with the predictions of density functional theory, but no temperature dependence through the magnetic transitions. Moreover, there is a number of surface-localized  features, with one of them having a metallic character. We were able to identify the dominant cleavage plane as the Al(1) layer through a combination of ARPES results, core-level photoemission data and slab calculations. However, the surface remains rough at the microscopic level with well-ordered Al(1)-terminated regions, alternating with regions where surface atoms are missing. Interestingly, certain fine features of the Fermi surface can only be explained after the inclusion of the magnetic moments of the \textit{f}-electrons. This means that the \textit{f}-electrons have an indirect effect on the electronic structure through their localized  spins even if their direct interaction with the conduction electrons remains small. As a matter of fact, low-dimensionality, surface-localized  features, dependence on the termination layer, exchange interaction and spin-orbit coupling are all present in the experimental electronic structure of TbAuAl$_4$Ge$_2$.  

\section*{acknowledgments}
The authors acknowledge SOLEIL and SOLARIS for the provision of synchrotron radiation facilities, proposals No. 20230207 (SOLEIL), 20227185 (SOLARIS) and 20232158 (SOLARIS). We would like to thank the beamline staff for their support during the experiment in beamline CASSIOPEE (SOLEIL), PHELIX (SOLARIS) and URANOS (SOLARIS). This publication was partially developed under the provision of the Polish Ministry and Higher Education project "Support for research and development with the use of research infra-structure of the National Synchrotron Radiation Centre SOLARIS” under contract no 1/SOL/2021/2. Work at ISMO was supported by public grants from the French National Research Agency (ANR), project MICROVAN No. ANR-24-CE30-3213-01, the ”Laboratoire d’Excellence Physique Atomes Lumi\`ere Mati\`ere” (LabEx PALM project MiniVAN) overseen by the ANR as part of the ”Investissements d’Avenir” program (reference: ANR-10-LABX-0039), and CNRS International Research Project EXCELSIOR. P.R.G. received financial support from the Brazilian agency CAPES under the process numbers 88887.642783/2021-00 and 88887.877620/2023-00. T.K. acknowledges Grant-in-Aid for Scientific Research (JSPS KAKENHI Grant No. JP23KJ0099) and support from GP-Spin at Tohoku University. R.B. and K.F. were supported by the National Science Foundation through Grant No. DMR-1904361.

\end{document}